\documentclass[aps,pra,twocolumn,showpacs,preprintnumbers,amsmath,amssymb,longbibliography]{revtex4-2}
\usepackage{soul}
\usepackage{amsmath}
\usepackage{esvect}

\usepackage{natbib}
\usepackage{graphicx}
\usepackage{color}
\usepackage{tabularx}
\usepackage{multirow}

\usepackage{wasysym}

\usepackage{amssymb}

\usepackage[utf8]{inputenc}

\usepackage{float}

\usepackage{subfigure}

\usepackage{float}

\usepackage{hyperref}

\usepackage{cancel}

\usepackage{xcolor}

\usepackage{ulem}

\usepackage{setspace}
\usepackage{soul}

\begin{document}

\title{Dual-frequency Doppler-free crossover resonance with suppressed magnetic-field sensitivity for compact optical frequency standards}

\author{D.\,S.~Chuchelov}
\author{M.\,I.~Vaskovskaya}
\author{E.\,A.~Tsygankov}
\email[]{tsygankov.e.a@yandex.ru}
\author{V.\,V.~Vassiliev}
\author{S.\,A.~Zibrov}
\author{V.\,L.~Velichansky}

\affiliation{P.\,N. Lebedev Physical Institute of the Russian Academy of Sciences,\\
Leninsky Prospect 53, Moscow, 119991 Russia}

\begin{abstract}
We report the observation and characterization of a high-contrast dual-frequency Doppler-free ground-state crossover resonance in the D$_1$ line of $^{87}$Rb. 
The crossover appears at a two-photon detuning exceeding the natural linewidth of the excited state and is formed by the optical pumping effect.
Unlike the previously proposed resonance at zero two-photon detuning, which we show becomes sensitive to magnetic-field fluctuations due to residual ellipticity of the optical fields—resulting in frequency shifts and profile asymmetry—the crossover resonance is largely immune to this effect. Our theoretical analysis attributes the observed sensitivity to the dispersive contribution of ground-state coherences to absorption.
Stability measurements under magnetic-field fluctuations demonstrate that using the crossover resonance provides more than an order-of-magnitude improvement, making it a promising reference for frequency stabilization in compact, field-deployable optical standards.
\end{abstract}

\maketitle
\section{Introduction}
The development of compact optical frequency standards is currently an area of active research, driven by~demands for high-precision and portable timekeeping systems~\cite{newman2019architecture}. 
A promising approach to achieving long-term frequency stability in such standards involves the use of~Doppler-free resonances~\cite{letokhov1977nonlinear,RevModPhys.54.697}, which can serve as reference for laser frequency stabilization.  
Currently, the most widely used reference is based on the two-photon transition $5$S$_{1/2}\rightarrow5$D$_{5/2}$ in $^{87}$Rb~\cite{PhysRevApplied.9.014019, PhysRevApplied.12.054063, Beard:24, 10722354, Duspayev_2024, obaze2025comprehensive}, which has shown excellent stability at the level of~$6\cdot10^{-14}$ at~$1$~s~\cite{ahern2025tailoring}. 
Another promising type of optical reference, relying on~dual-frequency sub-Doppler spectroscopy (DFSDS)  and observed in D$_1$ line of alkali-metal atoms, has attracted considerable attention in recent years~\cite{Hafiz:16, hafiz2017high, gusching2021short, gusching2023cs, DFDF87Rb}. 
This resonance arises from optical pumping mechanisms involving coherent population trapping (CPT) and nonlinear Hanle effects, under the action of counter-propagating bichromatic optical fields with orthogonal linear polarizations. 
While the resonance obtained using the DFSDS technique exhibits a greater width compared to the two-photon resonance, its advantages are a high amplitude and a less complex detection system.
 
A remarkable type of dual frequency Doppler-free resonance was previously predicted theoretically~\cite{hafiz2017high}.
This resonance occurs at two-photon (Raman) detuning greater than natural linewidth of the involved transitions, which takes it beyond the regime typically associated with DFSDS resonances. Similar features, referred to as inter-ground-state crossover resonances, were experimentally observed in the Cs D$_2$ line in~\cite{chang2021inter}.

In this work, we investigate the metrological performance of this ground-state crossover resonance in $^{87}$Rb D$_1$ line and propose it as a reference in optical frequency standards, since it as it exhibits
reduced sensitivity to variations in key experimental parameters, including magnetic field, two-photon detuning, cell position and counter-propagating field polarization compared to conventional DFSDS approach. 
Particular attention was paid to the influence of polarization ellipticity and the magnetic field on the lineshape asymmetry.
Our results show that this resonance is particularly well suited for use in compact frequency references, where environmental robustness and simplicity of implementation are essential.

\section{Experiment}
\label{SectionExperiment}

\subsection{Experimental setup}

\begin{figure}[t] 
\centering
\includegraphics[width=0.95\linewidth]{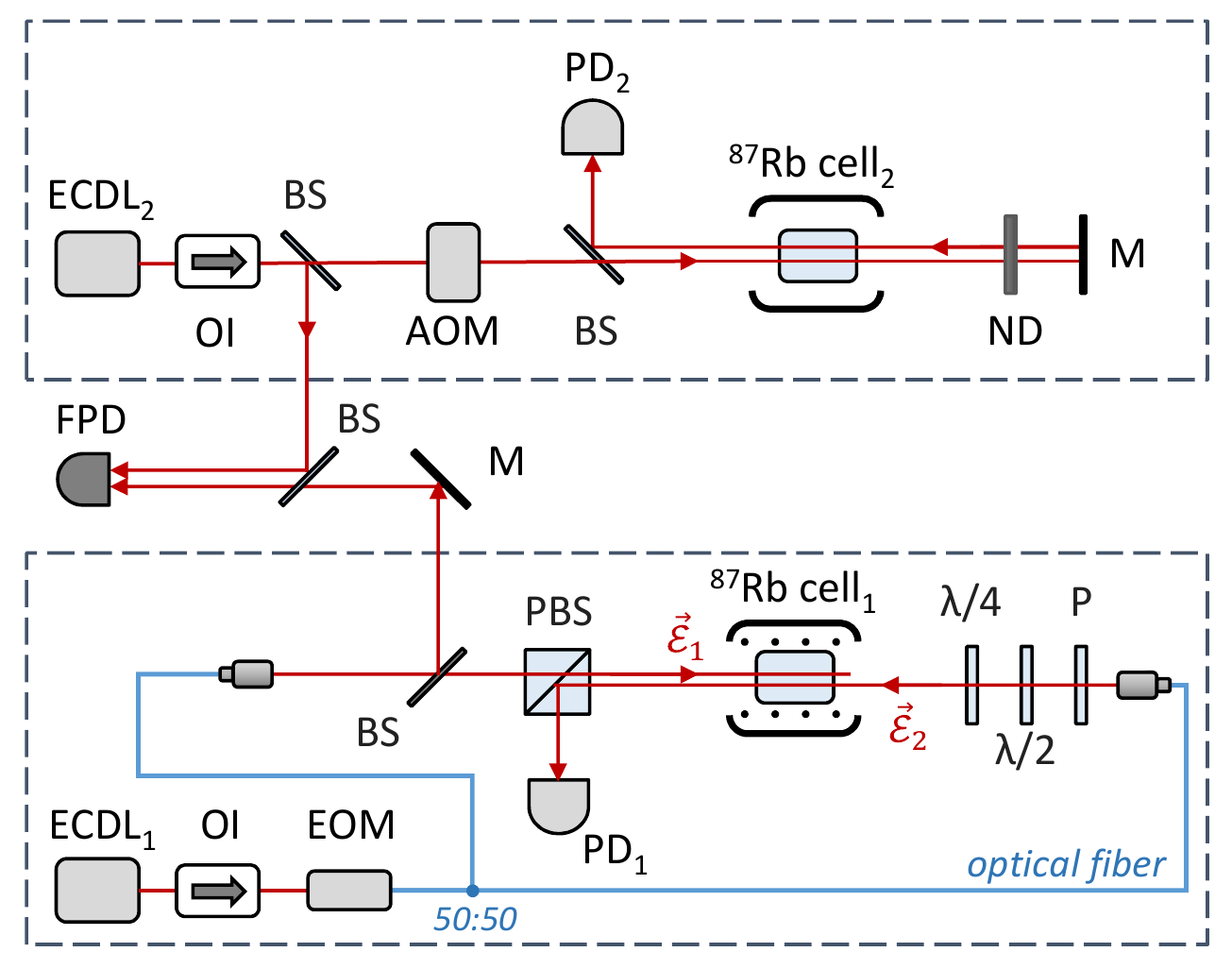}
  \caption{Schematic of the experimental setup. \hbox{ECDL---extended-cavity diode laser}, OI---optical isolator, EOM---electro-optic modulator, P---polarizer,  $\lambda/2$---half-wave plate, \hbox{$\lambda/4$---quarter-wave plate}, PBS---polarizing beam splitter, BS---beam splitter, \hbox{AOM---acousto-optic modulator}, ND---neutral density filter, M---mirror, PD---photodetector, FPD---fast photodetector. }
  \label{ExpSetupBypass}
\end{figure}

Fig.~\ref{ExpSetupBypass} shows the schematic of~the experimental setup. 
We used two custom-built external-cavity diode lasers (ECDL) operating at 795~nm, each incorporating an interference filter as the wavelength-selective element~\cite{vassiliev2019vibration}. 
The first system was designed to detect dual-frequency Doppler-free resonances.
The radiation of the ECDL$_1$ passed through an optical isolator and then through an electro-optic modulator (CAS-P-08-10-PP-FA) to generate the required bichromatic radiation.
The laser radiation was modulated at frequency $\Omega\approx3.4$~GHz, and the first-order sidebands of the spectrum were tuned to be resonant with the $F_g=1,2 \rightarrow F_e$ transitions of $^{87}$Rb D$_1$ line.
The modulation index was adjusted to achieve the highest amplitude of the first sidebands. 
The radiation at the output of the EOM was equally divided into two optical channels by a fiber splitter.
The pump field $\vec{\mathcal{E}}_1$ passed through a polarizing beam splitter and acquired horizontal linear polarization. 
A polarizer placed in the probe field $\vec{\mathcal{E}}_2$ path ensured its vertical polarization.
Both counter-propagating fields with orthogonal linear polarizations (lin$\perp$lin) were directed into the atomic cell.
Half- and quarter-wave plates were used to adjust the ellipticity of the probe field, with the fast axis of the quarter-wave plate aligned vertically.
The probe radiation transmitted through the cell was reflected by the polarizing beam splitter and registered by the photodetector.

The second system was used for traditional saturated absorption spectroscopy. The radiation of the ECDL$_2$ passed through an optical isolator, made a double pass through the atomic cell due to a retroreflecting mirror, and was registered by the photodetector. The return (probe) beam was attenuated by a neutral density filter.
Two identical cylindrical atomic cell, $20$~mm in diameter and $10$~mm in length, with AR-coated on four sides of the windows and filled with isotopically enriched $^{87}$Rb were used in experiment.
The cells temperature was maintained close to~$65$~$^\circ$C with an~accuracy of~$0.01$~$^\circ$C.
Each cell was placed inside a three-layer magnetic shield, and the first system included a solenoid that provided a longitudinal magnetic field.
A portion of the radiation from each laser system was split off and sent to~a~fast photodetector to~register the beat frequency.
An acousto-optic modulator driven by~$80$~MHz signal was employed in the second system to~shift the laser frequency.

\begin{figure}[t] 
\centering
\includegraphics[width=0.95\linewidth]{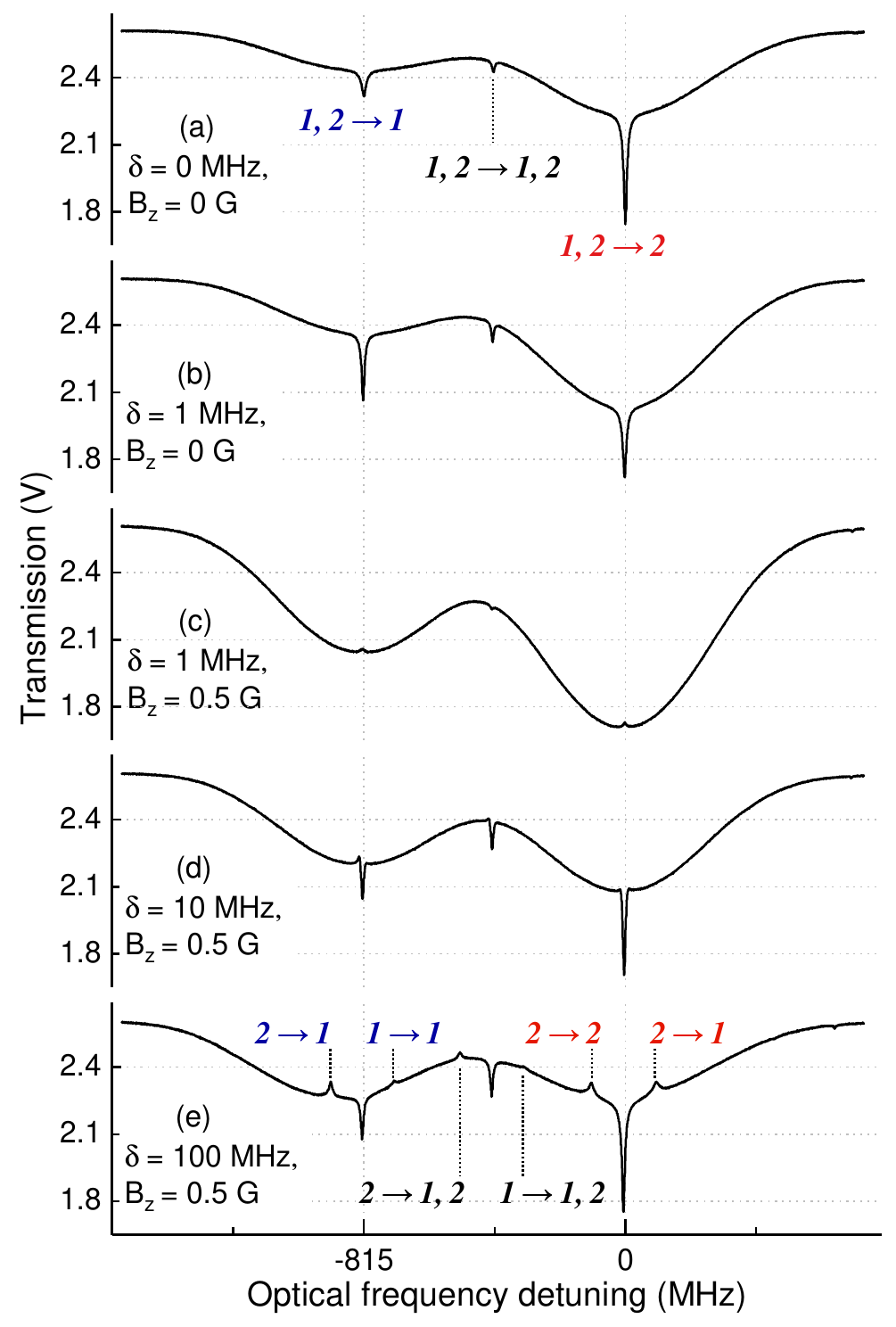}
  \caption{Dual-frequency Doppler-free spectra of~$^{87}$Rb D$_1$ line taken at~crossed polarizations of~the counter-propagating waves for different values of two-photon detuning~$\delta$  and magnetic field $B_z$. The horizontal axis represents the detuning of the laser carrier frequency from the value given by the sum of the $F_g=2 \rightarrow F_e=2$ transition frequency and half of the ground-state hyperfine splitting.}
  \label{Resonances}
\end{figure}

\subsection{Doppler-free spectra}

The Doppler-free spectra obtained in the dual frequency regime are shown in Fig.~\ref{Resonances} for different values of two-photon detuning $\delta=\nu_g/2 -\Omega$ ($\nu_g$ is the ground-state hyperfine splitting)
and longitudinal magnetic field $B_z$. 
Both counter-propagating waves had an optical power of $450~\mu\text{W}$ and a beam diameter ($1/e^2$) of approximately $2.5$~mm.
In the case of zero two-photon detuning and zero magnetic field, commonly referred to as the DFSDS approach, three  absorption peaks are observed (Fig~\ref{Resonances}\,a).
These inverted resonances are explained by~optical pumping processes into~hyperfine and Zeeman coherences associated with the CPT and Hanle effects, respectively.
At exact optical resonance, the counter-propagating waves interact with atoms whose velocity has zero $z$-component ($z$-axis taken along the propagation direction of $\vec{\mathcal{E}}_2$); 
however, due to the orthogonal linear polarizations of $\vec{\mathcal{E}}_1$ and $\vec{\mathcal{E}}_2$, dark states are not formed. 
In contrast, under nonresonant conditions, waves interact with atoms from different velocity groups, which ensures the formation of dark states and leads to~the suppression of absorption in the Doppler background. 
Among the three observed peaks, two correspond to resonances associated with transitions to one of excited sublevels, labeled as \hbox{$(1,2)\rightarrow1$} and \hbox{$(1,2)\rightarrow2$}. The third peak, labeled \hbox{$(1,2)\rightarrow(1,2)$}, is a crossover resonance.

Two-photon detuning influences both the amplitudes and widths of the resonances, as demonstrated in~\cite{DFDF87Rb} for detunings up to 1~MHz. Within this range, the amplitude of the high-frequency $(1,2)\rightarrow2$ peak decreases, whereas that of the low-frequency resonance increases (Fig~\ref{Resonances}\,b).
This effect arises from the destruction of hyperfine coherences, which leads to stronger absorption on the Doppler background for high-frequency resonance, while for low-frequency resonance the enhancement is more pronounced at the peak.
Upon applying the strong magnetic field Zeeman coherences are also destroyed leading to enhanced absorption across the entire Doppler profile and almost complete disappearance of Doppler-free resonances~(Fig.~\ref{Resonances}\,c).
As the two-photon detuning increases further, narrow dips reappear (Fig.~\ref{Resonances}\,d). 
When the value of $\delta$ exceeds the resonance linewidth, two small-amplitude ``side'' resonances can be resolved near each main peak, as shown in Fig.~\ref{Resonances}\,e. 
Importantly, under these conditions, the high-frequency resonance \hbox{$(1,2)\rightarrow 2$} exhibits an amplitude and width comparable to those observed at zero two-photon detuning and zero magnetic field.

\begin{figure}[b] 
\centering
  \includegraphics[width=0.95\linewidth]{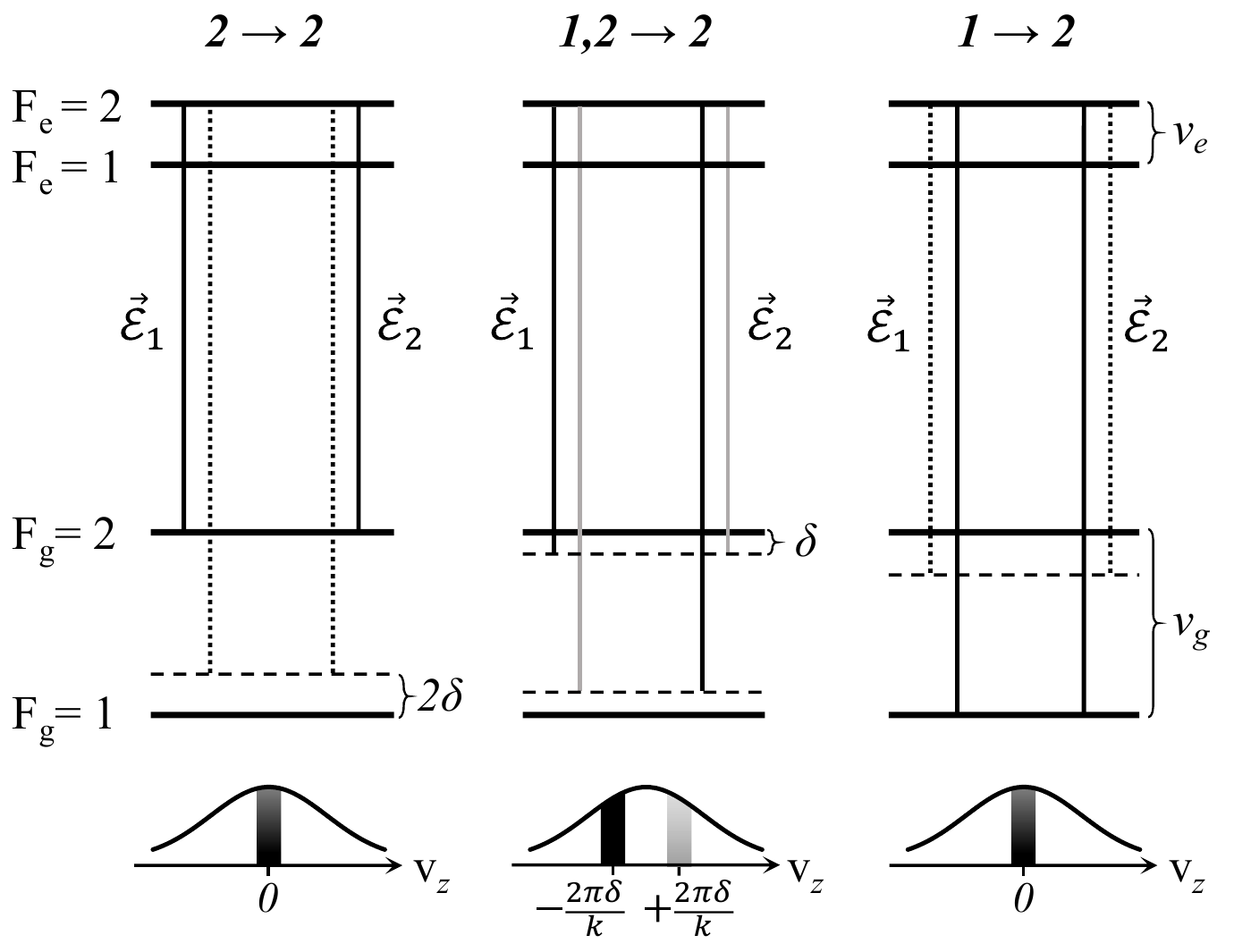}
  \caption{Schematic diagram of the energy levels for the D$_1$ line of $^{87}$Rb. The vertical lines represent first-order sidebands of counter-propagating fields for three laser carrier frequency detunings at a fixed two-photon detuning $\delta$. The lower part of the figure illustrates the Maxwell distribution of longitudinal atomic velocities and highlights the velocity groups that interact with the resonant optical components.}
  \label{ResonanceScheme}
\end{figure}

Below, we consider the formation of resonances corresponding to non-zero two-photon detuning using the example of a high-frequency resonance. 
We analyze the case where the two-photon detuning exceeds the natural width of transitions significantly, yet remains smaller than the Doppler broadened linewidth. 
Fig.~\ref{ResonanceScheme} shows energy levels and fields 
diagram which correspond to the peaks \hbox{$2\rightarrow2$}, \hbox{$(1,2)\rightarrow2$} and \hbox{$1\rightarrow2$}. 
The first of the listed peaks is observed when the low-frequency components of the counter-propagating bichromatic fields $\vec{\mathcal{E}}_1$ and $\vec{\mathcal{E}}_2$ are resonant with the transition $F_g=2\rightarrow F_e=2$ and interact with one group of atoms with a longitudinal velocity $v_z=0$. 
This situation is similar to conventional single-frequency saturated absorption spectroscopy. 
The transmission peak \hbox{$1\rightarrow2$} is formed in the same way, appearing when  the high-frequency components of the fields are resonant with the transition \hbox{$F_g=1\rightarrow F_e=2$}. 
The absorption peak \hbox{$(1,2)\rightarrow2$} arises when the counter-propagating fields interact with two velocity groups of~atoms with \hbox{$v_z=\pm2\pi\delta/k$}. 
The non-zero longitudinal velocity compensates for the two-photon detuning $\delta$, so~that the optical field components induce transitions from both ground-state sublevels within these atomic velocity group. 
For the atomic group with \hbox{$v_z = -2\pi\delta/k$}, the transition \hbox{$F_g = 2\rightarrow F_e = 2$} is driven by~the low-frequency component of the field $\vec{\mathcal{E}}_1$, while the transition \hbox{$F_g = 1 \rightarrow F_e = 2$} is driven by~the high-frequency component of the field $\vec{\mathcal{E}}_2$. 
For the atomic group with $v_z = +2\pi\delta/k$, the situation is reversed. 
In~this case, hyperfine optical pumping is suppressed, resulting in~a~significant increase in absorption.
The observed resonance can be considered a crossover; however, unlike the conventional crossover, it involves ground-state sublevels rather than those of the excited state.  

The low-frequency peaks \hbox{$2\rightarrow1$}, \hbox{$(1,2)\rightarrow1$}, and \hbox{$1\rightarrow1$} are formed as a result of a similar interaction between specific atomic velocity groups and the optical fields.
Three central peaks \hbox{$2\rightarrow(1,2)$}, \hbox{$(1,2)\rightarrow(1,2)$}, and \hbox{$1\rightarrow(1,2)$} in Fig.~\ref{Resonances}\,e are crossover resonances that  arise from the interaction of bichromatic radiation with atomic velocity groups involving both excited-state sublevels.
The \hbox{$2\rightarrow(1,2)$} and \hbox{$1\rightarrow(1,2)$} resonances correspond to saturated absorption spectroscopy crossovers, for which the interacting atoms have longitudinal velocities of $\pm\pi\nu_e/k$, where $\nu_e$ is the hyperfine splitting of the excited state. The $(1,2)\rightarrow(1,2)$ resonance includes four velocity groups with $v_z = \pm\pi(\nu_e - 2\delta)/k$ and $v_z = \pm\pi(\nu_e + 2\delta)/k$. For each group, two optical transitions are driven by the counter-propagating fields, involving all four sublevels.

\subsection{Influence of two-photon detuning and magnetic field on the resonance amplitude and width}
\begin{figure}[t]
   \centering
\includegraphics[width=0.95\linewidth]{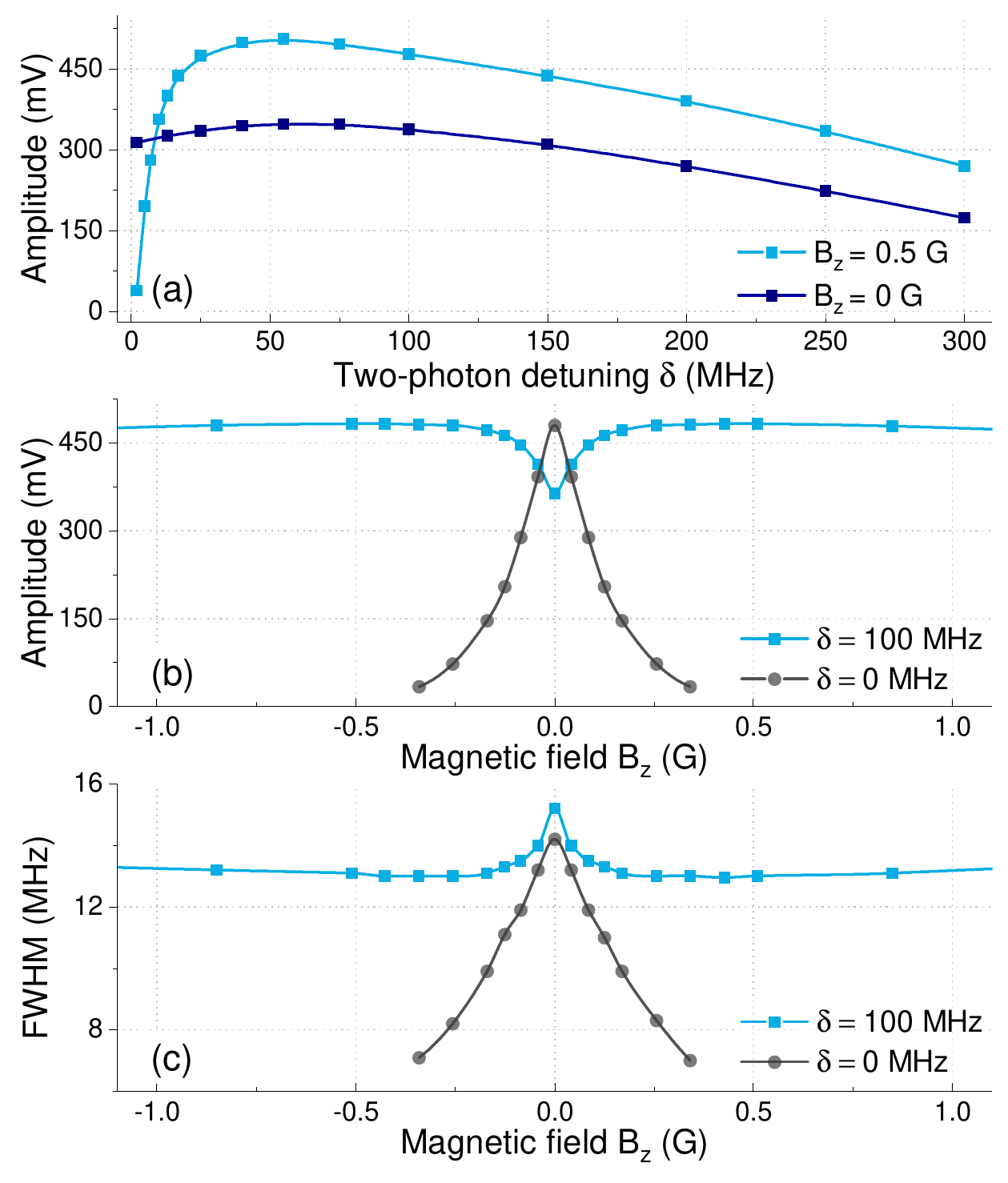}
    \caption{Dependence of the $(1,2)\rightarrow 2$ resonance amplitude on the two-photon detuning~(a). The first points correspond to Fig.~\ref{Resonances}\,c and ig.~\ref{Resonances}\,d.\\
    Dependence of the resonance amplitude~(b) and width~(c) on the longitudinal magnetic field.}
    \label{Ampldep}
\end{figure}

Fig.~\ref{Ampldep}\,a shows the amplitude of the high-frequency \hbox{$(1,2)\rightarrow\,2$} resonance as a function of the two-photon detuning $\delta$ for two magnetic field values. 
For zero field, the initial increase in amplitude is due to reduced absorption on the Doppler background caused by hyperfine optical pumping, while at larger detunings the amplitude decreases as fewer atoms interact resonantly with the fields. 
If the magnetic field is applied, the dependence is qualitatively similar, but the amplitude is~close to zero at~small detunings. 
In this case, Zeeman coherences  formed in both ground-state hyperfine sublevels are suppressed at the resonance peak as well as within the Doppler background.
However, due to strong hyperfine pumping, the contribution of Zeeman coherences to~absorption at~the background is smaller than at the peak. 
As a result, the amplitude of the Doppler-free resonance increases.

Fig.~\ref{Ampldep}\,b shows the dependence of the high-frequency resonance amplitude on the longitudinal magnetic field for two values of two-photon detuning.
At \hbox{$\delta=0$}, the resonance amplitude is maximal at zero magnetic field and drops down almost to zero as the magnetic field approaches the Earth's field (\hbox{$\approx0.5$~G}). This behavior is~explained by both the destruction of Zeeman coherences and the splitting of hyperfine coherences in the magnetic field, which results in increased absorption on the Doppler background.
In contrast, at large two-photon detuning, greater than the natural linewidth of the excited state, the resonance amplitude exhibits a minimum at~\hbox{$B=0$}.
The resonance amplitude grows with the magnetic field, reaching a maximum at approximately $0.4$~G, and then gradually decreases due to the splitting of the Zeeman structure. 
It can be seen that the maximal amplitude of the resonance is  nearly the same in both cases of $\delta$. 

Although the linewidth of the \hbox{$(1,2)\rightarrow2$} resonance~(Fig.~\ref{Ampldep}\,c) decreases with increasing magnetic field (for $|B_z|<0.5$~G) in both cases of $\delta$, their behavior differs significantly and follows a trend similar to that of~the amplitude.
The resonance width at $\delta=0$ decreases sharply with increasing magnetic field, which was also experimentally observed in~\cite{gusching2021short}, whereas for $\delta=100$~MHz, the width initially decreases slightly and then remains almost constant.
Hence, in the case of large two-photon detuning, there exists a broad range of magnetic field values ($0.2...0.5$~G) over which the sensitivity of the resonance characteristics to magnetic field fluctuations is reduced.

\begin{figure}[t]
    \centering
    \includegraphics[width=0.95\columnwidth]{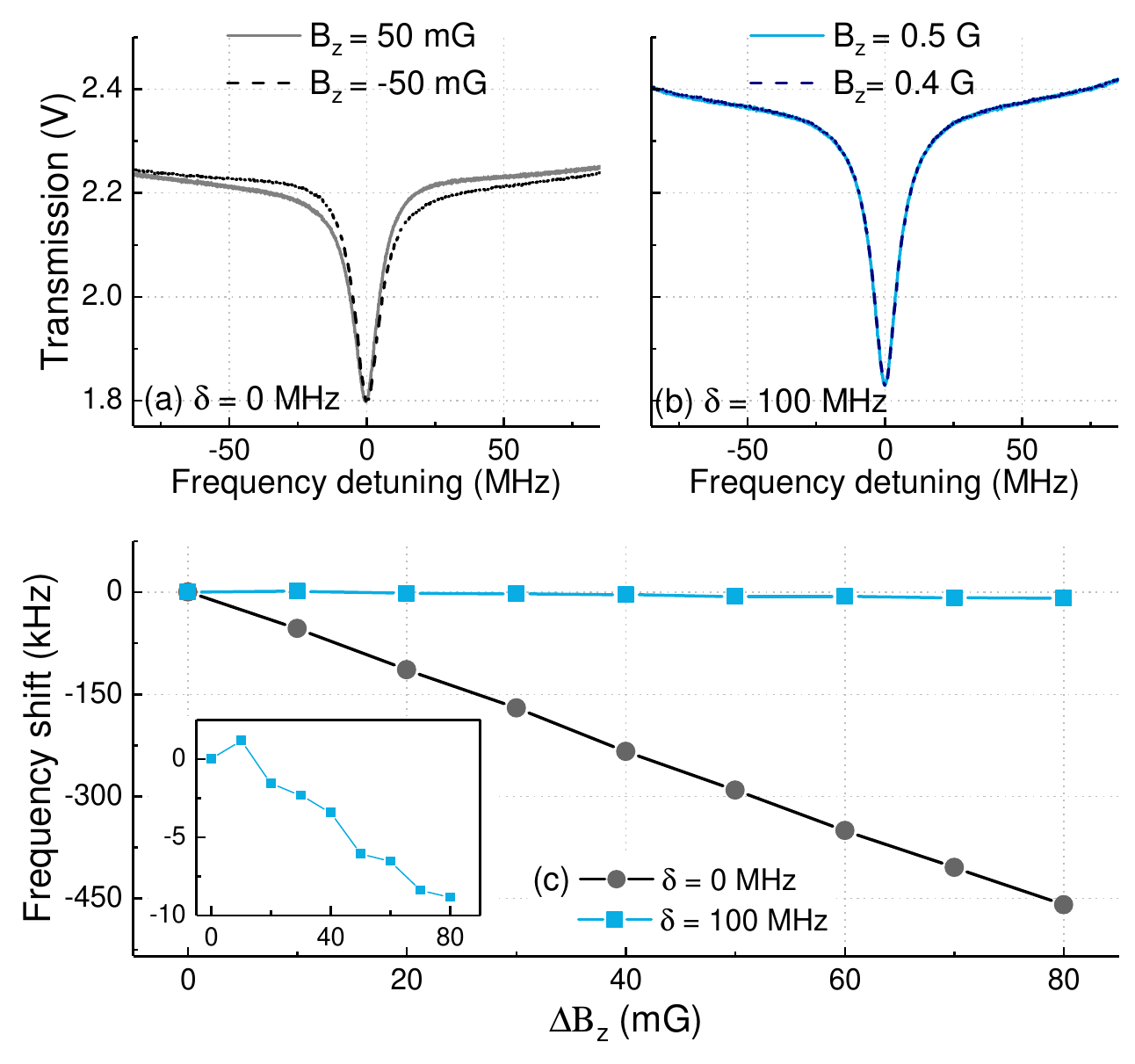}
    \caption{Doppler-free resonances for two values of longitudinal magnetic field $B_z$  in the case of elliptically polarized field $\vec{\mathcal{E}}_2$: $\delta=0$~MHz (a) and $\delta=100$~MHz (b). \\
    Dependences of the beatnote frequency shift on the change in the magnetic field magnitude $\Delta B_z$ in the case of elliptically polarized field $\vec{\mathcal{E}}_2$ (c). 
    The inset shows the blue curve with a rescaled vertical axis.}
    \label{Asymmetry}
\end{figure}

\subsection{Influence of polarization ellipticity on resonance asymmetry}

The asymmetry of the reference resonance profile is an important parameter that affects metrological characteristics. 
While the properties of the DFSDS resonance at zero two-photon detuning have been widely studied, this aspect has received little attention, despite the fact that number of experimental reports reveal pronounced asymmetry (for example, Figs.4,5 in \cite{hafiz2017high}).
We have analyzed the resonance profile under various experimental conditions and have found that it strongly influenced by polarization ellipticity, which may arise from imperfections of the optical elements or~their alignment. 
To investigate this effect, the polarization ellipticity of~the $\vec{\mathcal{E}}_2$ field was introduced by rotating the half-wave plate.
Fig.~\ref{Asymmetry}\,a and Fig.~\ref{Asymmetry}\,b present the  resonances for an elliptically polarized field $\vec{\mathcal{E}}_2$ under magnetic field variations of $\pm50$~mG for $\delta=0$ and $\delta=100$~MHz. 
The magnetic field was adjusted to satisfy the optimal conditions for each case.
The half-wave plate was rotated by~$5^\circ$, producing an ellipticity -- defined as the ratio of the intensities of the orthogonal linear components---of~approximately $0.03$, with the major axis of the polarization ellipse oriented vertically.
When the magnetic field is~applied the resonance at~$\delta=0$ shows a~pronounced asymmetry, which depends on the the magnetic field direction, while the resonance at~$\delta=100$~MHz remains nearly unaffected under the same conditions.
A qualitative explanation of the resonance asymmetry  at $\delta=0$  is as follows. 
The Doppler-free spectrum is determined by the Zeeman and hyperfine coherences associated with the Hanle and CPT resonances, respectively.
The lineshape of the Hanle resonance is of primary importance here.
At zero magnetic field, the absorption spectrum reflects the peak values of the Hanle resonance at different optical detunings.
Upon applying the magnetic field, the absorption corresponds to the position on the slope of the Hanle resonance.
For an elliptically polarized $\vec{\mathcal{E}}_2$ field, the Hanle resonance is asymmetric, with the ``sign'' and degree of the asymmetry depending on the detuning from the optical resonance. 
As a result, in a nonzero magnetic field, the Doppler-free resonance acquires an asymmetric lineshape due to unequal contributions from Zeeman coherences along its slopes.
For the case $\delta=100$~MHz, when a magnetic field is applied, such an effect is not observed due to the absence of Zeeman coherences.
A theoretical analysis of asymmetry is provided in the sections below.

To evaluate the effect of resonant asymmetry on its frequency, we~measured the beatnote frequency between two laser systems. 
The frequencies of two lasers were stabilized to~the Doppler-free resonances using standard procedures.
To~generate the error signal the  current of ECDL$_2$ was modulated at~frequency of~$30$~kHz. 
The frequency of ECDL$_1$ was modulated at $7$~kHz using a piezoceramic actuator to avoid possible effects of residual amplitude modulation.
The frequency deviation was set close to~the half-width of the reference resonance to optimize the steepness of the error signal.
After lock-in detection, the demodulated transmission signal was processed by a PID-controller, which generated a correction signal applied to the control inputs of both the piezoelectric transducer and the laser current in each laser system.
The beatnote signal of the two laser systems, detected by a fast photodiode, was analyzed using a frequency counter.
In the second laser system the~$+1$~diffraction order of the AOM was used for traditional saturated absorption spectroscopy, and the transition \hbox{$F_g = 2\rightarrow F_e = 2$} served as the reference.
In the first system the dual frequency Doppler-free resonance \hbox{$(1,2)\rightarrow2$} was used for frequency stabilization.

Fig.~\ref{Asymmetry}\,c shows the dependence of the beatnote frequency shift on the increment of the magnetic field magnitude for two values of the two-photon detuning with an elliptically polarized field $\vec{\mathcal{E}}_2$. Under these conditions, the resonance at $\delta=0$ is about $50$ times more sensitive to magnetic field variations than the resonance at~\hbox{$\delta=100$}~MHz. 
The lineshape asymmetry grows with both polarization ellipticity and magnetic field strength. The residual sensitivity of the resonance frequency at $\delta=100$~MHz is associated with optical pumping among magnetic sublevels by the elliptically polarized field and with Zeeman splitting in strong magnetic field.

\begin{figure}[b] 
\centering
  \includegraphics[width=0.95\linewidth]{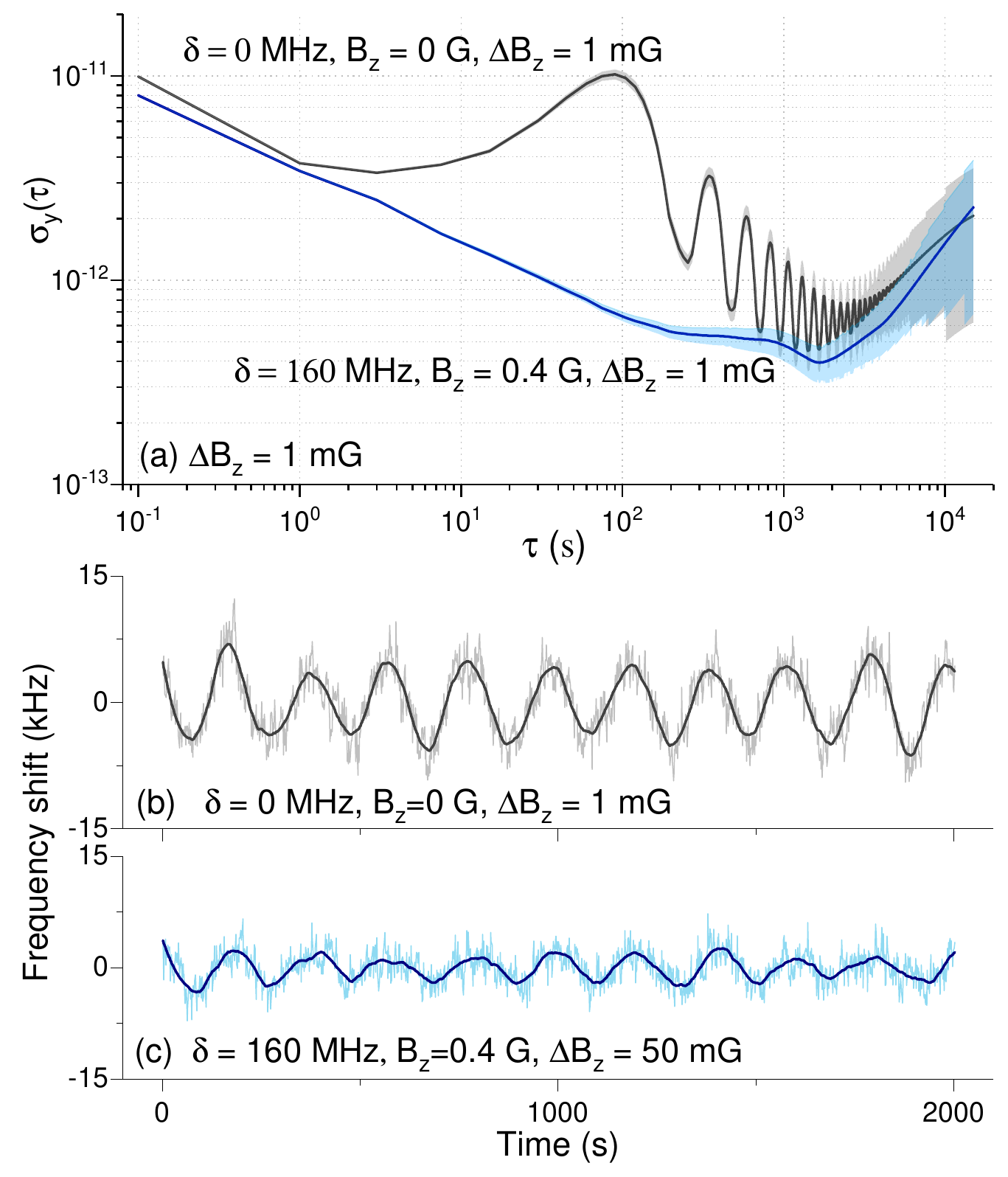}
  \caption{(a) Allan deviation of the laser beat frequency. A harmonic magnetic field modulation with amplitude $\Delta B_z$ and a period  $T=200$~seconds is applied. In (b) and (c), time-series measurement of the beatnote frequency between two laser systems are shown. The gate time is $1$~s.}
  \label{stability}
\end{figure}

\subsection{Influence of magnetic field fluctuations on~frequency stability}
We performed an experiment on frequency stability measurement to verify the  significance of our findings.  
The detection scheme of the sub-Doppler resonance in the first laser setup was modified to be more suitable for practical use in a compact device.
Instead of splitting the radiation into two optical channels, the counter-propagating lin$\perp$lin configuration was realized by employing a retroreflecting mirror and a double pass through a quarter-wave plate placed between the cell and the mirror.
In~the locked mode the $80$~MHz beatnote between the frequency of ECDL$_2$ and the first-order spectrum sidenband of~ECDL$_1$ generated by the EOM, was recorded at both zero two-photon detuning and at $\delta=160$~MHz.

Fig.~\ref{stability} presents the Allan deviation of the laser beatnote frequency for two different experimental conditions in first laser system. 
In both measurements the magnetic field was modulated harmonically with an amplitude 
$\Delta B_z= 1$~mG and a period of $T=200$~s. 
This level of~perturbation is comparable to that caused by the Earth’s magnetic field when the  device is rotated, assuming a~shielding factor of~$500$.
Under conditions corresponding to the conventional DFSDS regime ($\delta=0$, $B_z=0$) periodic frequency perturbations at the level of $10^{-11}$ were observed, manifesting as bumps in the Allan deviation plot at the half-period of the harmonic modulation. 
In~the second measurement, the ground-state crossover resonance was used ($\delta=160$~MHz, $B_z=0.4$~G), and no evident effect of magnetic field variation was observed in Allan deviation plot.
An increased modulation amplitude of 
$50$~mG  was applied to assess the sensitivity to magnetic field variations. Fig.~\ref{stability}\,b and Fig.~\ref{stability}\,c show
time-series measurement of the beatnote frequency between two laser systems in two regimes:~$\delta=0$~MHz with $B_z=0.001\sin(2\pi t/T)$~G, and $\delta=160$~MHz with $B_z=0.4+0.1\sin(2\pi t/T)$~G, respectively.
A reduction of about 50-times of magnitude in magnetic field sensitivity was achieved under our experimental conditions by employing the ground-state crossover resonance for frequency stabilization.
In~locked mode, a~frequency stability of~$4\cdot10^{-12}$ at~$1$~second was achieved in both cases, which was limited by~the laser's frequency noise at~the modulation frequency in first laser system.  The degradation of frequency stability at $\tau>1000$~s is primarily due to~temperature and laser intensity fluctuations.

\section{Theoretical analysis of the resonance asymmetry}
\label{SectionTheory}

\begin{figure}[t]
    \centering    \includegraphics[width=1
\columnwidth]{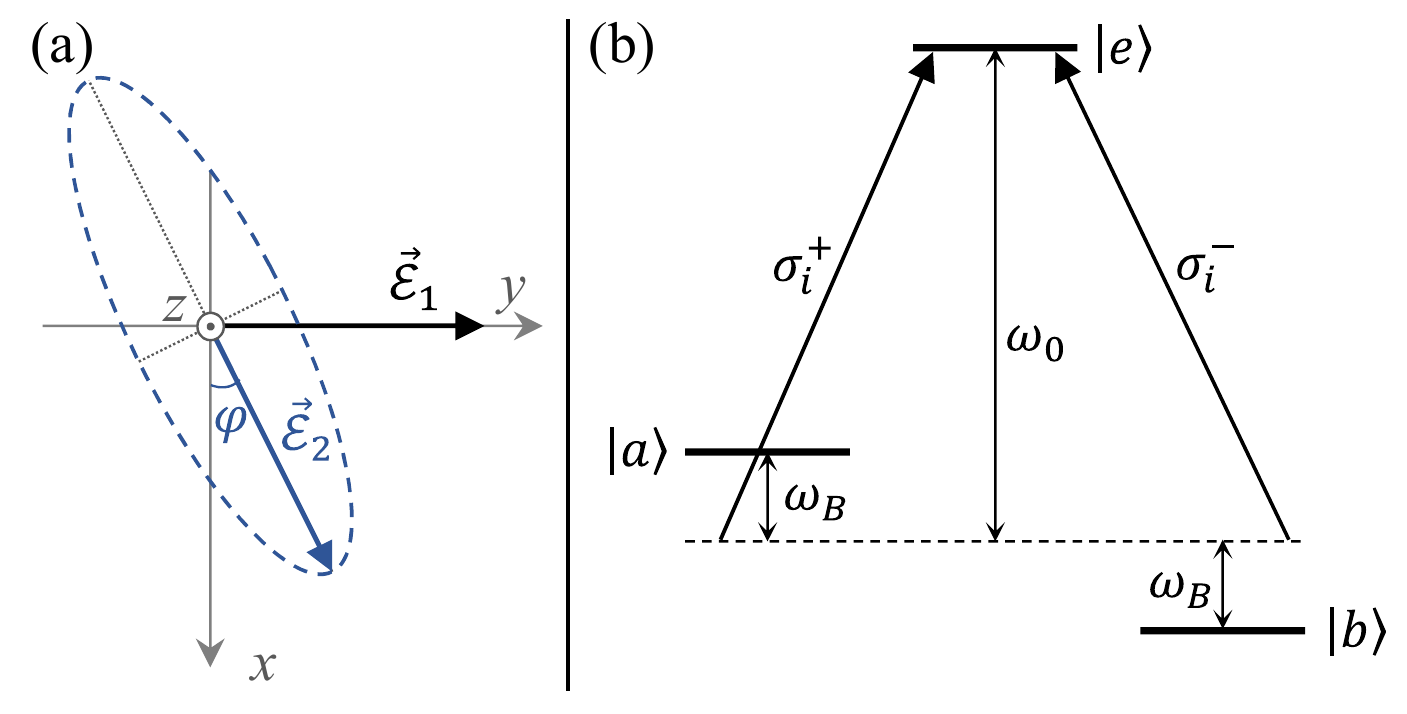}
    \caption{Configuration of the optical fields under consideration (a) and the energy levels diagram (b). The forward wave $\vec{\mathcal{E}}_1$ is linearly polarized, the backward wave $\vec{\mathcal{E}}_2$ is elliptically polarized. In~the theoretical model, both waves are decomposed into two components with orthogonal circular polarizations $\sigma_i^{\pm}$.}
    \label{LambdaDFDF}
\end{figure}

To describe the asymmetric Doppler-free resonance we~consider the $\Lambda$-system of~levels with lifted degeneracy of~the ground state due to~shift of~the levels $|a\rangle$ and $|b\rangle$ by~$\omega_{\mathcal{B}}$ in~the opposite directions; see Fig.~\ref{LambdaDFDF}. The ellipticity of~one of~the optical waves is~accounted~via unequal amplitudes $\mathcal{E}_-\neq\mathcal{E}_+$ of~its $\sigma^{\pm}$ components:

\begin{equation}
\begin{gathered}
\vec{\mathcal{E}}_2(t)=\mathcal{E}_+\dfrac{\vec{e}_x+i\vec{e}_y}{\sqrt{2}}\cos{\left(\omega_Lt-kz+\varphi\right)}\\
+\mathcal{E}_-\dfrac{\vec{e}_x-i\vec{e}_y}{\sqrt{2}}\cos{\left(\omega_Lt-kz-\varphi\right)}
\end{gathered}
\end{equation}

\noindent where $\vec{e}_{x,\,y}$ are the unit vectors, $\omega_L$ is~the field's frequency, $k$~is~the wave number and the phase $\varphi$ allows us~to~rotate the axis of~the polarization ellipse from $x$ to~$y$ by~this value. 
The other optical wave is~considered to~be~linearly polarized along the $y$ axis and represented as a superposition of orthogonal circular components with equal amplitudes $\mathcal{E}$.
The $\sigma^+$ component of~both fields induces electric-dipole transitions between $|a\rangle$ and $|e\rangle$, while the orthogonal component couples $|b\rangle$ to~$|e\rangle$.

The equations for the density matrix elements are treated in~the rotating wave approximation (\hbox{$\rho_{ea,\,eb}\rightarrow\rho_{ea,\,eb}e^{-i\omega_Lt}$}), the low saturation regime ($\rho_{ee}\ll\rho_{aa},\,\rho_{bb}$), and under adiabatic elimination of~the excited state. This yields:

\onecolumngrid
{\small
\begin{subequations}
\begin{equation}
\begin{gathered}
\rho_{ee}=\dfrac{V^2_+}{\left(\Delta_L-kv\right)^2+\gamma^2/4}\rho_{aa}+\dfrac{V^2_-}{\left(\Delta_L-kv\right)^2+\gamma^2/4}\rho_{bb}\\
+\dfrac{V_-V_+}{\left(\Delta_L-kv\right)^2+\gamma^2/4}\left(e^{-2i\varphi}\rho_{ab}+e^{2i\varphi}\rho_{ba}\right)+\dfrac{V^2}{\left(\Delta_L+kv\right)^2+\gamma^2/4}\left(\rho_{aa}+\rho_{bb}-\rho_{ab}-\rho_{ba}\right)
\end{gathered}
\label{rhoee}
\end{equation}
\begin{equation}
\begin{gathered}
i\dfrac{\partial}{\partial t}\rho_{aa}=-i\gamma\dfrac{V^2_+}{\left(\Delta_L-kv\right)^2+\gamma^2/4}\rho_{aa}+\dfrac{V_-V_+}{\left(\Delta_L-kv\right)^2+\gamma^2/4}\left[-i\dfrac{\gamma}{2}\left(e^{-2i\varphi}\rho_{ab}+e^{2i\varphi}\rho_{ba}\right)+\left(\Delta_L-kv\right)\left(e^{2i\varphi}\rho_{ba}-e^{-2i\varphi}\rho_{ab}\right)\right]\\
+\dfrac{V^2}{\left(\Delta_L+kv\right)^2+\gamma^2/4}\left[-i\gamma\rho_{aa}+i\dfrac{\gamma}{2}\left(\rho_{ab}+\rho_{ba}\right)-\left(\Delta_L+kv\right)\left(\rho_{ba}-\rho_{ab}\right)\right]+i\dfrac{\gamma}{2}\rho_{ee}
\end{gathered}
\end{equation}
\begin{equation}
\begin{gathered}
\left[i\dfrac{\partial}{\partial t}-2\omega_{\mathcal{B}}-\left(\dfrac{V^2_+}{\Delta_L-kv+i\gamma/2}-\dfrac{V^2_-}{\Delta_L-kv-i\gamma/2}\right)+i\gamma\dfrac{V^2}{\left(\Delta_L+kv\right)^2+\gamma^2/4}\right]\rho_{ab}=\\
-i\dfrac{\gamma}{2}\left(\rho_{aa}+\rho_{bb}\right)\left[e^{2i\varphi}\dfrac{V_-V_+}{\left(\Delta_L-kv\right)^2+\gamma^2/4}-\dfrac{V^2}{\left(\Delta_L+kv\right)^2+\gamma^2/4}\right]\\+\left(\rho_{bb}-\rho_{aa}\right)\left[\left(\Delta_L-kv\right)e^{2i\varphi}\dfrac{V_-V_+}{\left(\Delta_L-kv\right)^2+\gamma^2/4}-(\Delta_L+kv)\dfrac{V^2}{\left(\Delta_L+kv\right)^2+\gamma^2/4}\right].
\end{gathered}
\label{rhoab}
\end{equation}
\label{eqs}
\end{subequations}}
\twocolumngrid

Here $\gamma$ is~the natural width of~the excited state, $V_{\pm}=d\mathcal{E}_{\pm}/2\hbar$ and $V=d\mathcal{E}/2\hbar$ are the Rabi frequencies, $\Delta_L=\omega_L-\omega_0$ is~the optical field detuning, $kv$ is~the Doppler shift. The spontaneous decay of~the excited state equally populates levels $|a\rangle$ and $|b\rangle$ with the rate $\gamma/2$, i.e., no~branching is~assumed.

For simplicity of~calculations we~solve the system~\eqref{eqs} in~the tau approximation. The time derivative $\left(\partial/\partial t\right)\rho_{ab}$ is~replaced by~$\left(1/\tau\right)\rho_{ab}$, while $\left(1/\tau\right)\left(\rho_{aa}-1/2\right)$ is~substituted instead of~$\left(\partial/\partial t\right)\rho_{aa}$. The parameter $\tau$ can be~considered as~a~mean flight time of~the atoms through the optical beams. The population $\rho_{bb}$ is~replaced by~$1-\rho_{aa}$ due to~the conservation law $\text{Sp}\left(\hat{\rho}\right)=1$.

\begin{figure}[t]
    \centering
    \includegraphics[width=0.95\columnwidth]{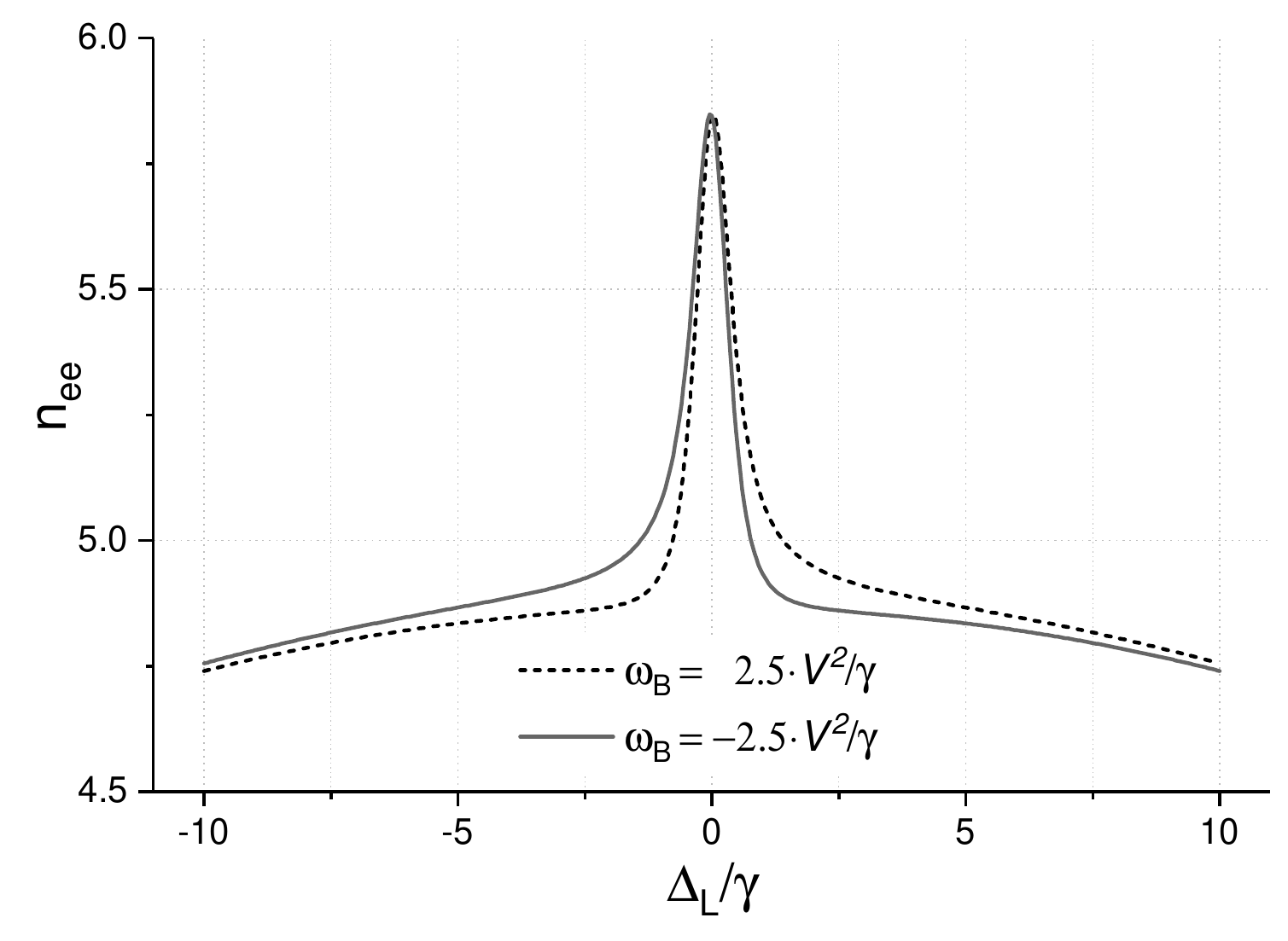}
    \caption{Calculated via~\eqref{eqs} Doppler-free resonance for elliptically polarized optical wave \hbox{$\left(V_+/V\right)^2=1.15$}, \hbox{$\left(V_-/V\right)^2=0.85$}. The solid curve is~plotted for $\varphi=\pi/16$, while the dashed curve is~given for $\varphi=-\pi/16$. The vertical axis is~given in~units $(V/\gamma)^2\cdot\gamma/\sqrt{\pi}kv_p$. The optical pumping factor $\tau V^2/\gamma$ is~equal to~$2/3$, $kv_p/\gamma=60$, which corresponds to~$^{87}$Rb atoms at~the temperature of~$300$~K, $v_p$~is~the most probable velocity. The curves were obtained by~integrating the excited-state population over the velocity with Maxwellian distribution.}
    \label{Asymmetrytheory}
\end{figure}

\begin{figure}[t]
    \centering
    \includegraphics[width=0.95\columnwidth]{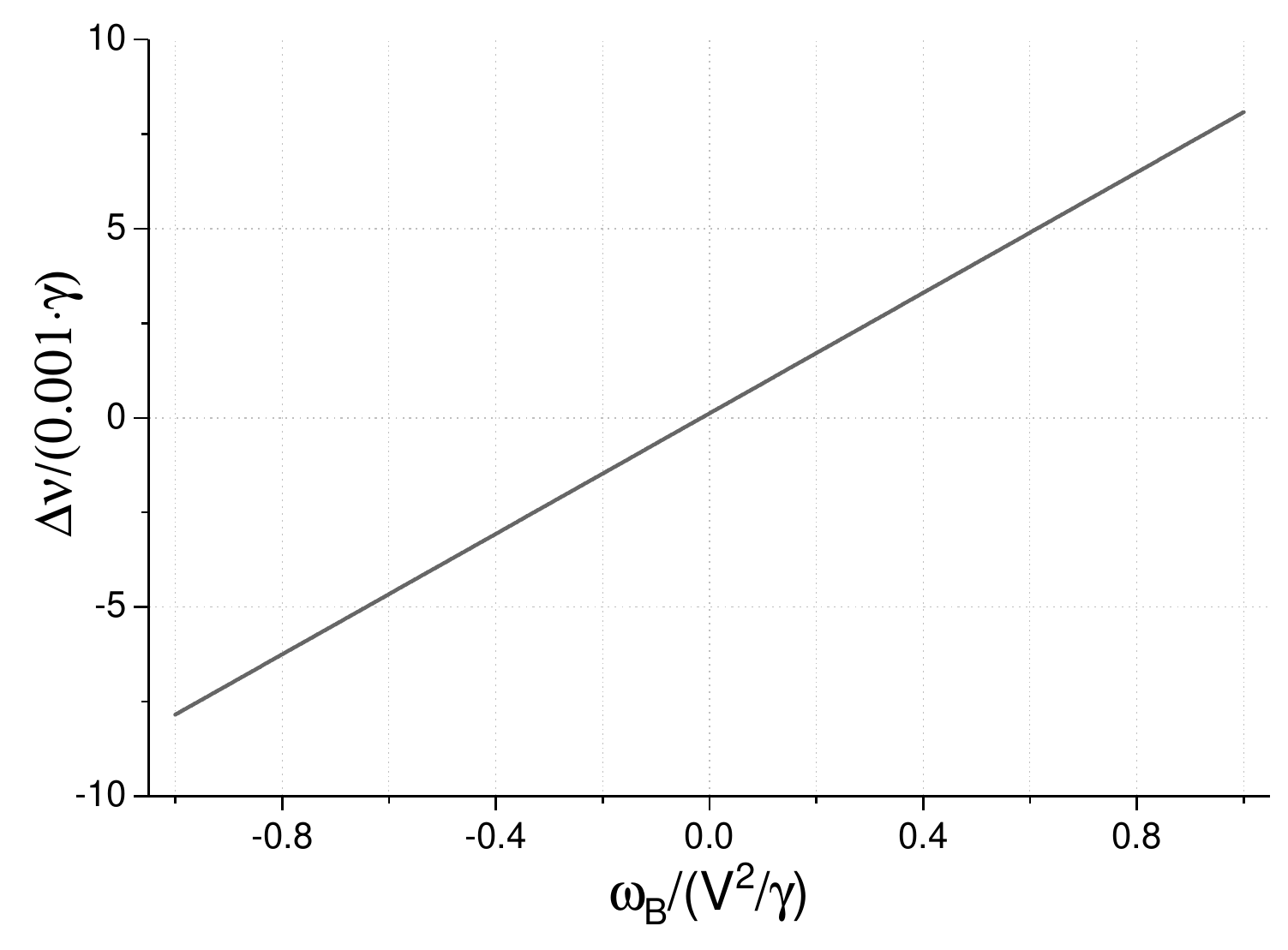}
    \caption{Calculated frequency shift of Doppler-free resonance for elliptically polarized optical wave \hbox{$\left(V_+/V\right)^2=1.15$}, \hbox{$\left(V_-/V\right)^2=0.85$}. The curve is~plotted for $\varphi=\pi/16$. The vertical axis is~given in units of $\gamma$. The optical pumping factor $\tau V^2/\gamma$ is~equal to~$2/3$.}
    \label{ShiftTheory}
\end{figure}

Fig.~\ref{Asymmetrytheory} demonstrates Doppler-free resonance for two opposite in~sign values of~$\omega_{\mathcal{B}}$ and $\varphi$, where shapes of~contours are different. To~understand the source of~the asymmetry, at~first see~Eq.~\eqref{rhoee} for excited-state population $\rho_{ee}$. It~contains terms \hbox{$\propto V_-V_+\text{Re}\left(e^{-2i\varphi}\rho_{ab}\right)$} and~\hbox{$\propto V^2\text{Re}\left(\rho_{ab}\right)$}. When $V_{-}\neq V_+$, the real part of~ground-state coherence~acquires odd~or~asymmetric over \hbox{$\Delta_L\pm kv$} contributions; see terms \hbox{$\propto(\rho_{bb}-\rho_{aa})$} in~Eq.~\eqref{rhoab}. They contribute to~the absorption at~$\omega_{\mathcal{B}}\neq0$ making~the Doppler-free resonance asymmetric. If~value of~$\omega_{\mathcal{B}}$ is~comparable to~width of~the ground-state transition, which is~determined by~the flight time and intensities of~the optical waves, there is~visible difference in~shapes of~the contours for the opposite values of~$\omega_{\mathcal{B}}$. Note also that terms $\propto\left(\Delta_L\pm kv\right)$ in~equations for the ground state drop linearly with this term when \hbox{$\left[\left(\Delta_L\pm kv\right)/\gamma\right]^2\gg1$}. Therefore, despite that $\omega_{\mathcal{B}}$ is~significantly smaller than $\gamma$, the asymmetric contribution to~the absorption has the width of~the optical scale. Also, at~nonzero values of~$\varphi$, absorption of~the elliptically polarized wave depends on~the imaginary part of~$\rho_{ab}$ induced by~the other optical field, which describes ground-state dispersion (odd over $\omega_{\mathcal{B}}$ function, if~we consider the Hanle resonance). And vice versa. This is~additional factor complicating the situation with asymmetry and frequency shift of~the resonance. When $\varphi\neq0$, the change in~$\omega_{\mathcal{B}}$ sign not only affects the resonance shape, but also changes the level of~absorption. To~obtain only the opposite asymmetry, the sign of~$\varphi$ should be~also changed. Fig.~\ref{Asymmetrytheory} demonstrates this property. The typical dependence of~the frequency shift on~the magnetic field is~given in~Fig.~\ref{ShiftTheory}. As~can be~seen, it~is linear in~the range of~moderate values of~$\omega_\mathcal{B}$. Note that the sign of the shift is opposite to the experimental one due to the choice of the corresponding sign of $\omega_B$ in the calculations.

\section{Summary}

High contrast dual-frequency Doppler-free resonances occur not only at zero two-photon detuning $\delta$ but also when it significantly exceeds the natural linewidth of the excited state. 
Such resonances observed in the D$_1$ line of $^{87}$Rb are identified as ground-state crossovers: only hyperfine optical pumping and Zeeman coherences are involved in their formation, whereas hyperfine coherences are not.
We characterize the metrological properties of both the conventional DFSDS resonance ($\delta$=0) and the crossover resonance as functions of two-photon detuning, magnetic-field magnitude, and the ellipticity of one optical field. 
The ground-state crossover exhibits high contrast and narrow linewidth, while showing significantly reduced sensitivity of these parameters to variations in magnetic-field strength.
Since hyperfine coherences do not contribute to the ground-state crossover, its amplitude, in contrast to the $\delta=0$ case, is insensitive to the position and size of the atomic vapor cell.
Moreover, the phase-noise level of~the microwave generator has no~impact on~the the ground-state crossover resonance characteristics.
Special attention was paid to the effects of polarization ellipticity and magnetic field on the asymmetry of the resonance contour and its central frequency.
The conventional DFSDS resonance exhibits strong magnetic sensitivity of its key characteristics and, under elliptical polarization, shows pronounced lineshape asymmetry and a frequency shift.
Theoretical analysis shows that this effect originates from the dispersive contribution of the ground-state coherences to the absorption.
Frequency-stability measurements under magnetic-field perturbations demonstrate that locking to the ground-state crossover yields more than an order-of-magnitude improvement in stability for $1$~mG field fluctuations---a perturbation level relevant for devices operating in the Earth’s field with a typical shielding factor of 500. 
These advantages make the ground-state crossover a promising reference for frequency stabilization in compact, field-deployable optical standards.

\section{Acknowledgments}
\label{Acknowledgments}

The authors receive funding from Russian Science Foundation (grant No. 24-72-10134).

\end{document}